\documentclass[aip, jcp, reprint]{revtex4-2}
\usepackage{amsmath}
\usepackage{xcolor}
\usepackage{graphicx}
\usepackage{mciteplus}
\usepackage{braket}
\usepackage[normalem]{ulem}
\usepackage{hyperref}

\usepackage[linesnumbered,lined,commentsnumbered,ruled]{algorithm2e}
\usepackage{rotating}
\usepackage{nicematrix}
\usepackage[T1]{fontenc}

\SetKwComment{Comment}{!! }{}

\newcommand{\rot}[1]{\rotatebox[origin=c]{90}{#1}}

\DeclareMathOperator*{\argmin}{arg\,min}

\begin{document}

\title{SOFI: Finding point group symmetries in atomic clusters as finding the set of degenerate solutions in a shape-matching problem}

\author{M. Gunde}
\email{miha.gunde@gmail.com}
\affiliation{Institute Ru\dj er Bo\v skovi\'c, Bijeni\v cka 54, 10000 Zagreb, Croatia}

\author{N. Salles}
\affiliation{CNR-IOM/Democritos National Simulation Center, Istituto Officina dei Materiali, 
c/o SISSA, via Bonomea 265, IT-34136 Trieste, Italy} 

\author{L. Grisanti}
\affiliation{Institute Ru\dj er Bo\v skovi\'c, Bijeni\v cka 54, 10000 Zagreb, Croatia}
\affiliation{CNR-IOM/Democritos National Simulation Center, Istituto Officina dei Materiali, 
c/o SISSA, via Bonomea 265, IT-34136 Trieste, Italy}

\author{L. Martin-Samos}
\affiliation{CNR-IOM/Democritos National Simulation Center, Istituto Officina dei Materiali, 
c/o SISSA, via Bonomea 265, IT-34136 Trieste, Italy} 

\author{A. Hemeryck}
\affiliation{LAAS-CNRS, Universit{\'e} de Toulouse, CNRS, 7 avenue du Colonel Roche, 31000 Toulouse, France}

\begin{abstract}
Point Group (PG) symmetries play a fundamental role in many aspects of theoretical chemistry and computational materials science. With the objective to automatize the search of PG symmetry operations of generic atomic clusters, we present a new algorithm called Symmetry Operation FInder (SOFI). SOFI addresses the problem of identifying PG symmetry by framing it as a degenerate shape-matching problem, where the multiple solutions correspond to distinct symmetry operations. The developed algorithm is compared against three other algorithms dedicated to PG identification, on a large set of atomic clusters. The results, along with some illustrative use cases, showcase the effectiveness of SOFI. The SOFI algorithm is released as part of the IRA library, accessible at \url{https://github.com/mammasmias/IterativeRotationsAssignments}.
\end{abstract}

\maketitle

\section*{NOTE}
This is the Accepted Manuscript version of the article. This article may be downloaded for personal use only. Any other use requires prior permission of the author and AIP Publishing. This article appeared in J. Chem. Phys. 161, 062503 (2024), and may be found at \url{https://doi.org/10.1063/5.0215689}.

\section{Introduction}

In the rise of quantum-chemistry and first-principle methods, the identification of Point Group (PG) symmetry has been a necessary step to effectively downsize the numerical problem and reduce the cost of simulations. 
Besides the
computational accelerations made possible by exploiting the symmetry,
the presence and identification of a PG symmetry in a molecule, or rather a cluster, or a defect, can be used to address the fundamental properties of its electronic ground, and excited states~\cite{gabi2022}.
Assigning symmetry label to each state (i.e. vibrational or electronic), is desirable to connect to group theory, which is the formal context where PG symmetries are described. 
If the symmetry is assigned to all relevant quantum (eigen)states, group theory can be applied to specific quantum-mechanical operators to understand which matrix elements are expected to vanish based on symmetry rules. 
This is extremely useful for example in molecular spectroscopy: depending on the operator governing the 
interaction term, the symmetry-based \textit{selection rules} will provide information on which transitions will actually be forbidden~\cite{Cotton,Atkins}.

As the identification of PG symmetry is not a new problem, the computational community has made numerous attempts to automatize it. 
Typically, quantum chemistry software and some molecular builders/editors include capabilities related to the symmetries of the system (i.e. automatic identification, symmetrization of structure). 
The algorithmic approaches to identify the PG may vary based on how the atomic structures are handled, and the way symmetry elements are utilised/exploited within the software framework. 
Despite the fundamental nature of PG symmetries and the broad use of related algorithms, there seem to be relatively few standalone libraries devoted to their identification.
Moreover, it is often very difficult to trace which PG identification algorithm (and/or library) is integrated into the larger software environment, as the information is often poorly, or not at all documented.
Many well-known quantum chemistry packages already date back to several years ago, when the modularity in programming was not necessarily a priority, so each developing group has created an "ad-hoc" method plugged into the code, often poorly updated through the years. 
For example by inspecting NWchem source code~\cite{nwchem2020} one can find a collection of old Fortran routines apparently written in the 90s, used for identifying symmetries. 
Besides, proprietary codes do not necessarily share information about their algorithm choice, even if some initiatives are more transparent, for example the IQmol builder~\cite{iqmol_link}, which uses, and references the SymMol~\cite{symmol} algorithm in its documentation.

In the subsequent paragraph we highlight a handful of approaches for the identification of PG symmetry, which have been devised as standalone algorithms/libraries, with some of them presently integrated into larger quantum chemistry codes, and/or computational "toolboxes".

The PG symmetry of a structure can be identified in several ways.
A relatively common strategy, used in several available approaches, is to first find the principal axis of symmetry, and then test for possible operations along other axes in relation to the principal axis.
The principal axes could be identified straight-forwardly by relying on the properties of the principal moments and axes of the inertia tensor, as done in SymMol~\cite{symmol}, and Symmetrizer~\cite{symmetrizer} software. 
Relying on the inertia tensor can however be problematic, since its principal axes can be ambiguous in some specific cases (spheroidal shapes).
Alternatively, certain properties of the atomic structure itself (atomic positions, midpoints between atoms, etc.) can be used to identify the axes of symmetry, such as employed in SYMMAPS~\cite{nedwed1994}, and SYVA~\cite{syva} software.
Another idea is to partition the atomic structure into sets of symmetry-equivalent atoms, and search for symmetries within those sets. 
The partitioning can be done through the inspection of the distance matrix of the structure, such as in libmsym~\cite{Johansson2017}, while the search for symmetries is conducted based on the properties of the inertia tensor of each equivalence set~\cite{beruski2014}, and the PG of the entire structure is the subset of the symmetry operations common to all partitions.
Some approaches based on arguments from graph theory also exist~\cite{ivanov1999, vandewiele2015}.

It is also common for PG-identification software to implement some other capabilities, such as precise symmetrization of structures~\cite{symmetrizer, syva}, or similar, which can be thought of as extensions to the fundamental task of identifying the PG symmetry elements of a structure.

In the interest of our discussion and the algorithm we propose, we now briefly recall some fundamental notions of PG symmetry. 
If $\boldsymbol{\boldsymbol{\theta}}$ is a specific symmetry operation, which is an element of the PG of an atomic structure $A$, the equation:
\begin{equation}
    A' \leftarrow \boldsymbol{\boldsymbol{\theta}} A,
    \label{eq:apply}
\end{equation}
represents the application of symmetry operation $\boldsymbol{\boldsymbol{\theta}}$ on structure $A$, which results in a symmetry-transformed structure $A'$.
By definition of symmetry operations, $A$ and $A'$ are equivalent up to a permutation of atomic indices $P_A$:
\begin{equation}
    A' = P_A A,
    \label{eq:permute}
\end{equation}
where $P_A$ is a permutation matrix, which is unique for a given operation $\boldsymbol{\boldsymbol{\theta}}$. Then, the relation:
\begin{equation}
    P_A A = \boldsymbol{\boldsymbol{\theta}} A.
    \label{eq:main}
\end{equation}
holds for any and every symmetry operation $\boldsymbol{\boldsymbol{\theta}}$ of the structure $A$. Each element (symmetry operation) of the PG of structure $A$ corresponds to a unique solution of Eq.~\eqref{eq:main}, in the form of a pair $\{\boldsymbol{\theta},P_A\}$. 

It can be observed that the Eq.~\eqref{eq:main} is equivalent to the shape matching problem of two exactly congruent~(equivalent) structures $A$ and $B$, as:
\begin{equation}
    P_B B = {\bf R} A + {\bf t},
    \label{eq:shapematching}
\end{equation}
where ${\bf R}$ is an orthonormal transformation matrix, and ${\bf t}$ is a translation vector.
By comparing Eq.~\eqref{eq:main} to Eq.~\eqref{eq:shapematching}, we see that ${\bf R}$ is replaced by a symmetry operation $\boldsymbol{\theta}$, and the translation ${\bf t}$ vanishes.
When solving the general shape matching problem of Eq.~\eqref{eq:shapematching}, only one solution (the optimal one) is sought, whereas in the case of PG identification in Eq.~\eqref{eq:main}, the set of all degenerate solutions represents the set of all possible symmetry operations of structure $A$.
Finding the PG of a structure can therefore be addressed as a "degenerate" shape matching problem.

In practice, the "degeneracy" of these solutions is never exact, due to numerical limitations, or distortions in the atomic positions within the structure.
An interpretation of "degeneracy" is thus needed, in terms of a threshold, such that all solutions within a given threshold are considered equivalent, degenerate.
By the tuning of such threshold, the notion of inexact symmetries can be covered.

In this work, we present the Symmetry Operation FInder (SOFI) algorithm, and its implementation as a standalone library, aimed at solving the general problem of PG symmetry identification, in a way that is independent of specific properties of the atomic structure.
SOFI finds the PG symmetry operations in the form of 3~$\times$~3 matrices, and the corresponding permutations of atomic indices. 
The working principle of SOFI is fundamentally based on ideas and arguments of the shape-matching algorithm IRA~\cite{ira,thesis}.
In contrast to other algorithms from the literature mentioned earlier, SOFI does not attempt to identify any special axes of symmetry, but is instead focused directly on generating the possible symmetry operations along {\it any} axis.
SOFI takes as input an atomic structure, and returns the list of symmetry operation matrices, written in the specific form of the original reference frame, in which the structure was input/written. 
As such, the matrices can directly be used without any post processing.
From this point of view, SOFI is a \textit{solver} for symmetry elements, that does not modify the input structure in any way.
In order to determine the PG of a structure, a separate routine is called, taking as input the list of symmetry operation matrices.
Within this routine, each symmetry operation is first associated to the axis it acts on, and labelled according to the Schoenflies notation~\cite{Atkins}. 
The PG is then deduced by inspection of the symmetry elements, following a standard flowchart.
The search for non-exact symmetries in SOFI can be tuned via a threshold parameter, whose value is directly linked to the distortions in the atomic positions of the structure. 
Moreover, in case of an incomplete list of elements of a PG, basic arguments of group theory can be used to find the missing elements.
An incomplete list of symmetry elements can be an indicator of broken symmetries in the system, which might happen when the atomic positions get distorted in a particular way.
Due to the modularity of the algorithm implementation, external conditions acting on symmetry elements can easily be implemented through custom functions. As an example of such function, the action of an external arbitrary-direction magnetic field on the PG of a structure~\cite{pausch2021} is included in the code.
SOFI is written in Fortran, with a C-bound Application Programming Interface (API), making it a directly interoperable library.
It can be used on its own, or easily inserted into other software.

Section~\ref{sec:algorithm} describes the details of our SOFI algorithm, and the Section~\ref{sec:in_practice} gives the implementation details.
In the Section~\ref{sec:benchmark}, SOFI algorithm is compared to three other algorithms for PG symmetry identification, namely SYVA~\cite{syva}, SymMol~\cite{symmol}, and libmsym~\cite{Johansson2017}.
The discussion in Section~\ref{sec:discussion} provides some possibly valuable insight into technical details of the algorithm, and the limits and tuning of its maximal "resolving" power.
The Section~\ref{sec:numeric} gives some exemplary use cases of the SOFI algorithm, and its output. 
Further examples can be found in the online documentation.

\section{The algorithm}
\label{sec:algorithm}
We approach the problem of finding PG symmetries by solving the degenerate shape-matching problem of Eq.~\eqref{eq:main}, but firstly we need to address the question of the origin point.

A fixed point is a point in space which remains unchanged under transformation by an arbitrary set of orthonormal matrices.
When an atomic structure is transformed by symmetry operations belonging to some PG, we can observe one or more such fixed points.
The points that remain fixed under transformation by all symmetry elements of the PG are the origin points of the entire PG, while points that are fixed only for a subset of the symmetry elements are origin points of the subgroups associated to that PG.
For example, the $Oh$ PG has six $C4v$ subgroups oriented in three different directions, each with a different origin point, while in the case of $C$-groups, all points along the primary rotation axis are fixed.
However, the only point of a generic structure, which is guaranteed to always remain fixed under transformation by any orthonormal matrix or symmetry element, is the geometric center (center of geometry, arithmetic mean, centroid).
Therefore, picking the geometric center of a structure as the origin point, guarantees that the algorithm will parse the full PG of the structure.
Indeed, most of the available algorithms assume this choice by default.
However, alternative origin points are possible.
For example when searching for symmetry operations about a given atom, the origin point should be the position of that specific atom (or some other specific point of choice).
SOFI is agnostic to the choice of origin, which means it will find symmetry operations about the given point.
This also means the origin has to be chosen from the outside (by the user, or another application).

After the origin point has been selected and the structure shifted accordingly, the atomic structure enters the main search loop of the SOFI algorithm.
The key idea of SOFI is to parse the space of orthonormal transformations, at each step construct a trial transformation, evaluate Eq.~\eqref{eq:main} with the trial transformation, and keep track of all favourable solutions.
Thus, the algorithm needs two main components, as follows.

The first component is a function which will evaluate Eq.~\eqref{eq:main}. The idea employed in SOFI is to transform Eq.~\eqref{eq:main} into a distance $d( P_A A, \boldsymbol{\theta} A)$, which should be equal to zero when $P_A$ and $\boldsymbol{\theta}$ correspond to a solution of Eq.~\eqref{eq:main}, and greater than zero otherwise.
In order to compute the distance $d(P_A A, \boldsymbol{\theta} A)$, we employ the Hausdorff distance function~\cite{Eiter1997, ira, thesis} with the generic form:
\begin{equation}
    d_H(A, B) = \max_{a\in A} \min_{b\in B} d(a,b)
    \label{eq:dH}
\end{equation}
where $A$ and $B$ are structures with atomic vectors $a\in A$, $b\in B$, and $d(a,b)$ is a Cartesian distance.
The Hausdorff distance is invariant to permutations, and variant to other rigid transformations of either structure (rotation, reflection, translation).
The evaluation of Eq.~\eqref{eq:main} is thus done 
by computing Eq.~\eqref{eq:compute_D}.
\begin{equation}
    d(P_A A,\boldsymbol{\theta} A)~=~d_H(A,\boldsymbol{\theta} A)    
    \label{eq:compute_D}
\end{equation}
If $\boldsymbol{\theta}$ is a solution of Eq.~\eqref{eq:main}, then $d_H(A,\boldsymbol{\theta} A) = 0$. 
In order to compute $d_H$ in SOFI, we use the CShDA algorithm, introduced in Ref.~\cite{ira}. 
CShDA imposes a one-to-one assignment constraint on the atoms, and returns the permutation $P_A$ and distance $d_H$.

The second component of the algorithm is a way of parsing the space of orthonormal transformations (3~$\times$~3 orthonormal matrices $\boldsymbol{\theta}$). 
This parsing should be computationally efficient, but still broad enough to identify all possible solutions of Eq.~\eqref{eq:main}.
In SOFI, we define the parsing of transformation space in a similar way as in IRA~\cite{ira}, in which the orthonormal transformations (bases) are constructed directly from atomic vectors present in the structure.
It follows from Eq.~\eqref{eq:main}, that for each atom $i\in A$ with the atomic position $a_i$, there exists a unique atom $j\in A$ with the atomic position $a_j$, such that:
\begin{equation}
    a_i = \boldsymbol{\theta} a_j.
    \label{eq:cob}
\end{equation}
The Eq.~\eqref{eq:cob} can be seen as operation of change of basis, from the basis $\boldsymbol{\beta}$ defined by $a_i$, to basis $\boldsymbol{\gamma}$ defined by $a_j$, written:
\begin{equation}
    \boldsymbol{\beta} a_i = \boldsymbol{\gamma} a_j,
    \label{eq:bas}
\end{equation}
from which we see that 
\begin{equation}
    \boldsymbol{\theta} = \boldsymbol{\beta}^T\boldsymbol{\gamma}.
    \label{eq:betaTgamma}
\end{equation}
Therefore, the operation $\boldsymbol{\theta}$ is completely defined by two orthonormal bases, namely $\boldsymbol{\beta}$ and $\boldsymbol{\gamma}$, which are defined by atomic positions $a_i$ and $a_j$, respectively.

However, an orthonormal basis set in 3D cannot be uniquely defined from a single 3D vector. 
This problem is resolved by adding another atomic vector $a_{i'}$ to the definition of the basis, as follows. 
The first basis vector $\hat{e}_1$ is the normalised atomic vector $a_i$, the second basis vector $\hat{e}_2$ is the component of the atomic vector $a_{i'}$ orthonormal to $\hat{e}_1$, and the third basis vector is a cross product $\hat{e}_3 = \hat{e}_1 \times \hat{e}_2$.
In this way, the basis $\boldsymbol{\beta}$ is defined by positions of atoms $i$ and ${i'}$, and the basis $\boldsymbol{\gamma}$ by positions of atoms $j$ and ${j'}$.
Due to this, SOFI is restricted to structures containing at least 3 non-collinear atoms. 
The same procedure for constructing an orthonormal basis is used in IRA~\cite{ira}.

According to the preceding considerations, in order to identify any 
symmetry operation $\boldsymbol{\theta}$, we need to find the quadruplet of atoms $\{i, {i'}, j, {j'} \}$, which construct $\boldsymbol{\theta}$ by Eq.~\eqref{eq:betaTgamma}, such that Eq.~\eqref{eq:main} is satisfied. 
In order to find all possible solutions of Eq.~\eqref{eq:main}, we need in turn to list 
all unique matrices $\boldsymbol{\theta}$ that can be constructed by such quadruplets of atoms, while also satisfying $d_H(A,\boldsymbol{\theta} A)=0$ by Eq.~\eqref{eq:dH}.
The most naive algorithm to find all symmetry operations $\boldsymbol{\theta}$ is then to brute-force search over all combinations of quadruplets of atoms in $A$, which give $d_H =0$.  
Such algorithm is constructed by four nested DO loops over all atoms, meaning an $\mathcal{O}(N^4)$ complexity, where $N$ is the number of atoms in the structure.
We can however quickly observe that such approach contains a huge degree of redundancy, which can be greatly reduced, or even avoided, by applying the following considerations.

\subsection{Reduce complexity I: fix one basis}
\label{sec:fix_one}
The Eq.~\eqref{eq:cob} holds true for a given $\boldsymbol{\theta}$, for all atoms $i$ simultaneously. 
Therefore, instead of looping over all $i$, ${i'}$, $j$, and ${j'}$, we can simply select any two different atoms $i$ and ${i'}$, and loop over only $j$ and ${j'}$.
The atoms $i$ and ${i'}$ construct the basis $\boldsymbol{\beta}$, which is kept fixed throughout the search, while the basis $\boldsymbol{\gamma}$ constructed by atoms $j$ and ${j'}$ keeps changing, effectively constructing different trial $\boldsymbol{\theta}$ operations.
The argument also holds for all possible operations $\boldsymbol{\theta}$, since each true symmetry operation $\boldsymbol{\theta}$ is associated to a single permutation $P_A$.
Thus the complexity of the search algorithm is reduced to $\mathcal{O}(N^2)$.

\subsection{Reduce complexity II: preserve distances and atomic types}
\label{sec:preserve_dist}
A PG symmetry operation $\boldsymbol{\theta}$ is a unitary transformation, meaning it preserves distances. 
Thus, an atom after the transformation cannot have a position which is at a different distance from the origin than it was before the transformation. 
Moreover, upon transformation by $\boldsymbol{\theta}$, the position of atom $i$ is effectively replaced by the atom $j$, which must have the same properties as atom $i$.
Hence atom $j$ should possess the same atomic type, and be positioned at the same distance from the origin, as atom $i$. 
This can be translated into a simple constraint for the choice of atom $j$ -- and also ${j'}$ -- namely the norms of atomic vectors (distances from the origin) must be equal, $||a_i||=||a_j||$ and $||a_{i'}||=||a_{j'}||$; and the atomic types of the pairs $\{i, j\}$ and $\{{i'},{j'}\}$ must be identical (see lines 4,5 and 7,8 of Alg.~\ref{alg:sofi_loop}).
Thus the complexity of the search algorithm is further reduced, to $\mathcal{O}(N_i N_{i'})$, where $N_i$ and $N_{i'}$ are the number of atoms within the same radial distance and atomic type as atom $i$, and atom ${i'}$, respectively.

\subsection{Reduce complexity III: pick atoms close to origin}
\label{sec:sort_atms}
The numbers $N_i$ and $N_{i'}$ are further reduced by choosing $i$ and ${i'}$ which are as close as possible to the origin (avoiding the atoms at the origin, for numerical reasons), and are of the atomic type which has the least number of equivalent atoms in that radial shell.
For close-packed structures (clusters), the worst-case scenario is a structure of a single atomic type, where $N_i$ and $N_{i'}$ are 12 (hexagonal close-packed single-crystal structure), which means 144 possible constructions of $\boldsymbol{\theta}$ by the SOFI algorithm.
For some types of structures however, these numbers can be a lot higher,
the extreme case being fullerene-type structures where all the $N$ atoms are at the same distance from the origin, $N_i=N_{i'}=N$. 
For those structures, the use of SOFI as described and implemented in the present paper is not efficient.

\subsection{A word about (roto-)reflections}
\label{sec:flip_ax}
Notice that when the bases $\boldsymbol{\beta}$ and $\boldsymbol{\gamma}$ are generated with the procedure defined earlier, the matrix $\boldsymbol{\theta}$ resulting from Eq.~\eqref{eq:betaTgamma} is always such that its determinant is positive, $det(\boldsymbol{\theta})=1$. 
This means our search algorithm always produces trial transformations which correspond strictly to rotations.
In order to mitigate this behaviour, we systematically flip each of the three vectors for each basis $\boldsymbol{\gamma}$ that is generated (see line 11 of Alg.~\ref{alg:sofi_loop}).
As such, the resulting trial transformations include proper rotations, reflections, and roto-reflections.
This slightly raises the complexity of the algorithm by a factor of 4 (one for the original $\boldsymbol{\gamma}$, and additional three for each flipped axis).

\subsection{Inexact symmetries}
\label{sec:inexact_symm}

A structure can exhibit inexact symmetries due to disorder or inaccuracies in atomic positions, which can be purely numerical, or due to physical reasons. In any case, this disorder can cause some symmetry elements to become inexact, \textit{i.e.} the distance $d_H(A, \boldsymbol{\theta} A)$ does not reach zero, but only a local minimum. For this reason we introduce the threshold $\texttt{sym\_thr}$ into SOFI, which modifies the condition for a transformation $\boldsymbol{\theta}$ to be recognised as symmetry operation of a structure $A$ into $d_H(A, \boldsymbol{\theta} A) \le \texttt{sym\_thr}$, which is interpreted as the maximal deviation (in units of distance) between an atom of the non-transformed structure, and its associated atom of the transformed structure, must be below the threshold.
The value of \verb|sym_thr| is an input parameter to SOFI.
For the same reason of possible disorder in the atomic structure, the transformations $\boldsymbol{\theta}$ which are constructed from atomic vectors by Eq.~\eqref{eq:betaTgamma}, can potentially become inexact (in the sense of non-zero $d_H$). In order to obtain as precise transformation matrices $\boldsymbol{\theta}$ as possible, we introduce the concept of matrix refinement, see Sec.~\ref{sec:in_practice} and Alg.~\ref{alg:refine}.

It can still happen that SOFI's parsing procedure will not identify all the symmetry operations of a structure due to such atomic disorder. In that case, arguments of fundamental group theory can be used to generate the missing symmetry operations. 
The most important argument is the property of group closure, which states that if $\boldsymbol{\theta}_i \in G$ and $\boldsymbol{\theta}_j \in G$ then also the product $\boldsymbol{\theta}_i*\boldsymbol{\theta}_j \in G$, where $G$ is the point group.
Therefore, after the main parsing procedure of SOFI is performed, 
the list of found symmetry operations can be looped through itself, and all possible products of the operations generated, then tested for the symmetry condition on the structure, and finally included in the set accordingly. Thus if any symmetry operation has been missed by the main search algorithm for any reason, it will be found and added by this (optional) procedure. 

The value given to the threshold $\texttt{sym\_thr}$ is directly linked to the level of disorder in atomic positions on the structure.
Its extremal value can ultimately be related to the distance between nearest atoms in the structure, in the sense that if the value $\texttt{sym\_thr}$ is close to, or larger than that, the algorithm will accept operations that effectively exchange the positions of atoms, as valid symmetry operations. 
In order to avoid that, the rule of thumb is that $\texttt{sym\_thr}$ can sensibly be set to any values up to about half the distance between nearest atoms, anything beyond that should be carefully tested, and will most probably produce results that will be hard to interpret.

\section{Implementation}
\label{sec:in_practice}

The main loop of the SOFI search procedure is schematised in Alg.~\ref{alg:sofi_loop}, which follows the considerations from Sec.~\ref{sec:algorithm}. 
Each iteration of the main loop consists of constructing a new orthonormal basis $\boldsymbol{\gamma}$, and with it a new trial operation $\boldsymbol{\theta}'$. 
Each trial $\boldsymbol{\theta}'$ is passed to the routine $\texttt{try\_sofi}$ (schematised in Alg.~\ref{alg:try_sofi}). 
This routine first determines if the matrix is already a known symmetry operation, and if not, computes the first $d_H(A,\boldsymbol{\theta}' A)$ via the CShDA algorithm. 
If the first $d_H$ is small enough, the trial $\boldsymbol{\theta}'$ is considered as potential symmetry operation, and passed to the $\texttt{refine}$ routine (schematised in Alg.~\ref{alg:refine}). 
The $\texttt{refine}$ routine returns a refined trial $\boldsymbol{\theta}'$ and the associated $d_H(A,\boldsymbol{\theta}'A)$. 
If this distance $d_H$ is below the threshold $\texttt{sym\_thr}$, the trial $\boldsymbol{\theta}'$ is stored as a new symmetry operation of $A$. 
At the end of the main loop, the list of symmetry operations $\boldsymbol{\theta}$ of structure $A$ is known.

\begin{algorithm}[h]
    \KwIn{structure $A$, threshold $\texttt{sym\_thr}$.}
    \KwOut{List of symmetry operations $\boldsymbol{\theta}$.}
    \BlankLine
    Select $i$ and $i'$ close to origin\Comment*[r]{Sec~\ref{sec:sort_atms}}
    $\boldsymbol{\beta} \gets$ OrthonormalBasis( $i$, ${i'}$)\Comment*[r]{Sec~\ref{sec:fix_one}}
    \For{$j \in 1, N$}{
       ${\bf if}$( $abs(|a_i| - |a_j|) > \texttt{sym\_thr}$ )${\bf cycle}$\Comment*[r]{Sec~\ref{sec:preserve_dist}}
       ${\bf if}$( $typ(i) \ne typ(j)$ )${\bf cycle}$\;
       \For{$j' \in 1, N$}{
          ${\bf if}$( $abs(|a_{i'}| - |a_{j'}|) > \texttt{sym\_thr}$ )${\bf cycle}$\;
          ${\bf if}$( $typ(i') \ne typ(j')$ )${\bf cycle}$\;
          $\boldsymbol{\gamma} \gets$ OrthonormalBasis( $j$, ${j'}$ )\;
          \For{$m\in 0,3$}{
             flip axis $m$ of $\boldsymbol{\gamma}$\Comment*[r]{Sec~\ref{sec:flip_ax}}
             $\boldsymbol{\theta}' = \boldsymbol{\beta}^T \boldsymbol{\gamma}$;
             {\bf call} $\texttt{try\_sofi}$($A$, $\boldsymbol{\theta}'$, $\texttt{sym\_thr}$)\;
          }
       }
    }
   \caption{The main search loop of SOFI, including the considerations from Sec.~\ref{sec:fix_one}-\ref{sec:flip_ax}. The called routine $\texttt{try\_sofi}$ is schematised in Alg.~\ref{alg:try_sofi}.}
   \label{alg:sofi_loop}
\end{algorithm}

\begin{algorithm}[h]
    \KwIn{structure $A$, 3$\times$3 matrix $\boldsymbol{\theta}'$, threshold $\texttt{sym\_thr}$.}
    \KwOut{Updated list of symmetry operations $\boldsymbol{\theta}$.}
    \BlankLine
    Compare $\boldsymbol{\theta}'$ to current list of all $\boldsymbol{\theta}$\;
    ${\bf if}$( $\boldsymbol{\theta}' \in$ list )${\bf return}$\;
    Compute $d = d_H(A, \boldsymbol{\theta}'A)$ via CShDA algorithm\;
    ${\bf if}$( $d > 5.0*\texttt{sym\_thr}$ )${\bf return}$\;
    {\bf call} $\texttt{refine}(A,\;\boldsymbol{\theta}',\;d_H)$\;
    ${\bf if}$( $d_H \le \texttt{sym\_thr}$ )${\bf add\;} \boldsymbol{\theta}' {\bf \;to \;list}$\;

   \caption{The schema of routine $\texttt{try\_sofi}$. The called routine $\texttt{refine}$ is schematised in Alg.~\ref{alg:refine}.}
   \label{alg:try_sofi}
\end{algorithm}

A trial matrix transformation $\boldsymbol{\theta}'$, whose action on a structure $A$ is sufficiently close to some true symmetry operation $\boldsymbol{\theta}$, can be identified, and \textit{refined} to correspond to the true symmetry $\boldsymbol{\theta}$, see Alg.~\ref{alg:refine}.
The matrix $\boldsymbol{\theta}'$ is said to be sufficiently close to $\boldsymbol{\theta}$, when the distance $d_H(A, \boldsymbol{\theta}' A)$ is below a given parameter value, but not zero (this value in SOFI corresponds to $5 *\texttt{sym\_thr}$, see line 4 of Alg.~\ref{alg:try_sofi}). 
In this case, the permutation of atoms $P_A$ that is returned when computing $d_H(A,\boldsymbol{\theta}' A)$ with CShDA, is not exact, but can be said to be partially correct, meaning that there are at least some atoms for which the assignments are correct.
This partially correct assignment can be applied to the structure $A$, and used as input to the well-known SVD-based algorithm described in Ref.~\cite{arun_svd}, which returns an optimal rotation given a fixed permutation of atoms.
The small rotation $\nu$ which is found by SVD is added to the original transformation, $\boldsymbol{\theta}''=\nu\boldsymbol{\theta}'$, and applied to the structure, $\boldsymbol{\theta}'' A$. 
At this point, the distance $d_H(A, \boldsymbol{\theta}'' A)$ and the corresponding assignments $P_A''$ are recomputed.
If $P_A''$ contains permutations, the cycle of SVD-CShDA is repeated.
Once the new assignment does not contain any further permutation, the procedure is stopped, the refined matrix is $\boldsymbol{\theta}''=\nu\boldsymbol{\theta}'$, and if $d_H(A,\boldsymbol{\theta}'' A) \le $ \texttt{sym\_thr}, then the refined matrix $\boldsymbol{\theta}''$ is a true symmetry operation $\boldsymbol{\theta}$.
The process is schematised in Algorithm~\ref{alg:refine}.
\begin{algorithm}[h]
    \KwIn{structure $A$, trial matrix $\boldsymbol{\theta}'$.}
    \KwOut{refined matrix $\boldsymbol{\theta}'$, distance $d_H$.}
    \BlankLine
    $\boldsymbol{\theta}''  \gets \boldsymbol{\theta}'$\;
    $P_A, d_H \gets CShDA( A, \boldsymbol{\theta}'' A )$\;
    \For { $n \in 1, \mathrm{max\_iter}$ }{
      $\nu \gets SVDrotation( A, \boldsymbol{\theta}'' A)$\;
      $\boldsymbol{\theta}'' \gets \nu\boldsymbol{\theta}''$\;
      $P_A, d_H \gets CShDA( A, \boldsymbol{\theta}'' A )$\;
      \textbf{if} (no permutation in $P_A$): exit loop\;
     }
     \Return $\boldsymbol{\theta}' \gets \boldsymbol{\theta}''$, $d_H$
    \caption{The schema of routine $\texttt{refine}$.}
   \label{alg:refine}
\end{algorithm}

The self-sufficient iteration scheme of Algorithm~\ref{alg:refine} is similar to the algorithm Iterative Closest Point (ICP)~\cite{besl}, which is very well known in the computer vision community. 
ICP has some known pitfalls, and several proposed solutions to them \cite{pottmann2006}. 
The most notable being getting stuck in local minima, or oscillating between two equivalent solutions.
It should be stated that these pitfalls can in principle occur in our matrix refinement approach as well, however, they occur in situations when $d_H(A,\boldsymbol{\theta}' A)$ is far from zero, and above any meaningful threshold, thus the trial transformation $\boldsymbol{\theta}'$ in such situations is far from a suitable candidate for our purposes, and it is not of concern.
Moreover, the number of iterations in Algorithm~\ref{alg:refine} has a hard limit at \texttt{max\_iter=3}.

In order to determine the PG from a list of symmetry operations in the form of 3 $\times$ 3 matrices, as returned by SOFI, we have implemented a decision process based on a flowchart available in physical chemistry textbooks \cite{Atkins}.
The decision tree is based on the identification and classification of specific symmetry elements.
For our purpose, the simplest way of carrying it out is to employ the Schoenflies notation for the symmetry elements, in the form of $Op ~n^{\wedge} p$, where $Op$ is the name of the operation ($C$ for rotation, $S$ for rotoreflection, $I$ for inversion, and $E$ for identity), $n$ is the order of rotation, and $p$ the multiple of rotation, for example $S~12^{\wedge} 5$ corresponds to a rotoreflection with an angle of $\frac{5}{12} 2\pi$.
The analysis of a 3 $\times$ 3 matrix, and assignment of its Schoenflies label is done as described in the Appendix~\ref{app:matrices}.
Once the list of symmetry operations is converted to the Schoenflies notation, and grouped together by the axis they act on, the decision tree to determine the PG is applied.

The SOFI software library is organized as a series of independent routines. 
For performing a default (full) search for PG of a structure, there is a wrapper routine which includes calls to all the needed routines. 
This includes shifting the structure to its geometric center before entering SOFI, and using the closure property of the groups to fully identify all symmetry operations (see Sec.~\ref{sec:inexact_symm}).
A workflow for more specialized tasks can be constructed as-desired from the routines included in the library.

The source code repository also includes an API to the main SOFI routines, which is bound to the C-namespace via the $\texttt{iso\_c\_binding}$ intrinsic Fortran module.
The API is compiled into a dynamic library (file $\texttt{.so}$), and thus the interface routines can easily be linked and called from either Fortran or C, or imported into Python.
Example programs for each of these languages are included in the repository. 
An appropriate Python module is also proposed, which acts as an interface to the C-bound routines in the dynamic library, through the $\texttt{ctypes}$ Python module.

\section{Comparison to other algorithms}
\label{sec:benchmark}

The SOFI algorithm has been compared to the SYVA~\cite{syva}, SymMol~\cite{symmol}, and libmsym~\cite{Johansson2017} software.
The full details and results of the tests are available in the Supplementary Materials, and the associated Zenodo repository~\cite{tests_zenodo} which contains all the structures, software, and scripts used in the test, and all results in plain text format.
The goal of this comparison is to showcase the behaviour of SOFI in comparison to other established software.

All of the atomic structures included in our comparison test were obtained from the Cambridge Cluster Database (CCD)~\cite{cdb}. 
They include atomic cluster structures, where we have excluded any structures with number of atoms less than 3.
The total number of structures included in the test is thus 3858, with the number of atoms varying between 3 and 380. 
There is only one set of structures with more than one atomic type, which has two atomic types.

The comparison tests were carried out by computing the PG of each input structure, at different values of the symmetry threshold. 
Changing the value of this threshold can result in the respective algorithm  detecting a different number of possible symmetry operations, and hence a different assignment of the PG.
The SYVA and SymMol algorithms have a single adjustable parameter for the tolerance, which, to our understanding, is more or less equivalent to the symmetry threshold \verb|sym_thr| used in SOFI.
That is, it specifies the maximal allowed displacement of any atom of the symmetry-transformed structure, with respect to the original structure.
Therefore, it is possible to directly compare the results of SYVA, SymMol, and SOFI algorithms at the same value of tolerance.
On the other hand, the libmsym algorithm does not have a single-valued parameter for the tolerance, but rather a set of parameters which act upon different segments of the algorithm, and it is therefore not straight-forward to compare the results.
Any computation where a respective algorithm outputs an explicit message of error or failure, or the software crashes with no output, is recorded as a failure.

The values for the tolerance used in SYVA, SymMol, and SOFI were $tol=0.001$ and $tol=0.05$, while for libmsym we used three sets of parameters which are also used in the Avogadro software~\cite{avogadro2012}, where they are labelled as "default", "medium", and "loose".

To the best of our knowledge in the CCD database, the source/algorithm which was used for obtaining the listed PGs is not clearly specified. 
This is an issue since the claimed PG cannot be actually tested or revised.
In other words, the PG assignment by CCD cannot be assumed to have a benchmarking value.

\begin{figure}
    \centering
    \includegraphics[width=\linewidth]{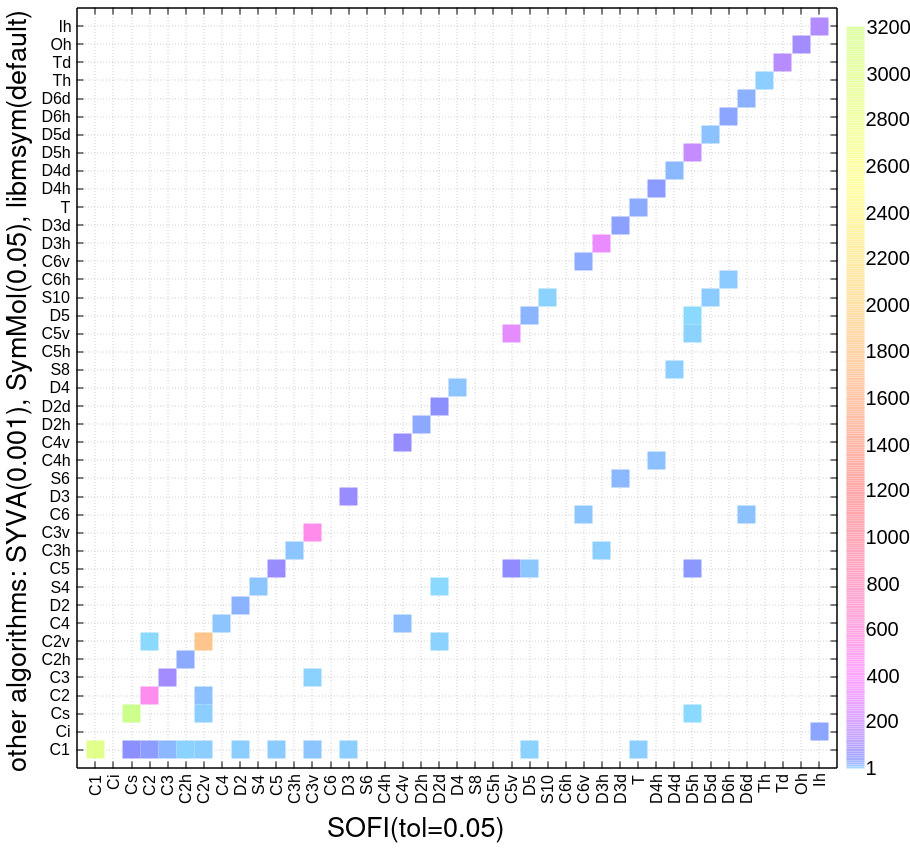}
    \caption{
    Compact summary of the comparison test results: 
    The horizontal axis represents PGs as assigned by SOFI, and the vertical axis PGs assigned by any other algorithms in the test (SYVA, SymMol, libmsym). 
    Labels of PGs are sorted by the cardinality of the groups. 
    The points below the diagonal represent cases where SOFI assigned a PG with higher order, and values above the diagonal represent cases where SOFI assigned a PG with lower order. The color of each point represents the number of structures where that particular combination of PG assignment happened.
    }
    \label{fig:rc_sofi_all}
\end{figure}

The Fig.~\ref{fig:rc_sofi_all} gives a correlation plot of the PGs as assigned by the different algorithms included in the test, excluding CCD.
This plot has been made by combining the results of each algorithm in the test, using the value of tolerance for which that algorithm produced the least failures. 
These are explicitly SYVA($tol=0.001$), SymMol($tol=0.05$), libmsym( $tol="default"$), and SOFI($tol=0.05$).
A point on the diagonal of the plot represents PG assignments of SOFI equal to any of the other algorithms, while an off-diagonal point represents different assignment of PGs, for the same structure. 
The distance of a point from the diagonal could be seen as proportional to the extent of the difference in the assignment, in terms of number of symmetry operations of a group.
Similar plots of this type are available in the Supplementary Materials, resolved per each algorithm and tolerance value included in the test.

We show in the Supplementary Materials, that SOFI is fully capable of computing the PG symmetry in atomic clusters.
Contrary to other algorithms in the test, there are zero cases of failure in SOFI, among all the structures and tolerance values included in the test.
In all cases of disagreement between SOFI and another algorithm, SOFI is capable of finding a PG of the same order, or higher, when computed with an appropriate tolerance value.

The small differences in SYVA, SymMol, and SOFI most probably arise due to the different ways in which algorithms find the symmetry operations, or their axes.
A small difference in the axis on which a symmetry element acts, can lead to a small difference in the atomic positions of the symmetry-transformed structure, which leads to a different value of distance to the original structure.
When such small discrepancies lead to distance values that are comparable to the value of symmetry threshold used, they can be sufficient to cause an algorithm to detect PGs of a different (lower) order.
These small variations can be observed in our tests, where in some cases SOFI detects less symmetry operations than SYVA and SymMol, however upon slight increase in the threshold, the higher-order PGs are always retrieved.

A more in-depth analysis of all the results would require a detailed inspection of the participating algorithms, and a case-to-case analysis of each result, which is well beyond the scope of the present article.
Besides, our tests did not investigate the impact of the symmetry threshold in a systematic way, which makes drawing any general conclusions from this comparison test nontrivial.

The Dzugutov-mod clusters dataset was used to give some indicative timing of the algorithms. 
It contains 248 structures with number of atoms from 3 to 250, all of the same atomic type.
The structures where an algorithm was not able to determine a PG were excluded from this measurement, since they take a disproportionate amount of time.
The SYVA algorithm took 155.8~s to determine 248 PGs, SymMol took 2.1~s to determine 245 PGs, libmsym took 0.8~s to determine 245 PGs, and SOFI took 1.2~s for 248 PGs.
All the timing measurements were done on single Intel Core i7-10700K processor.

\section{Discussion}
\label{sec:discussion}

Due to the SVD-based "matrix refinement" procedure introduced in SOFI (Sec.~\ref{sec:in_practice}, Alg.~\ref{alg:refine}), the axis of a symmetry element is ultimately set such that the root-mean-square distance between the original, and the symmetry-transformed structure $d_{rmsd}(A,\boldsymbol{\theta}A)$ is minimized.
However it should be noted that, in general, the value of $\boldsymbol{\theta}$ which minimizes a distance function $D$, in the sense of Eq.~\eqref{eq:argmin},
\begin{equation}
    \argmin_{\bf \boldsymbol{\theta}} \big\{ D({\boldsymbol{\theta}}A, A)\big\}     
    \label{eq:argmin}
\end{equation}
can be slightly different for different distance functions $D$, such as the root-mean-square $d_{rmsd}$, and Hausdorff distance $d_H$, the latter of which is used to evaluate the operations $\boldsymbol{\theta}$ in SOFI by Eq.~\eqref{eq:compute_D}.
This is a small inconsistency in the SOFI algorithm, and it means the value of Eq.~\eqref{eq:compute_D} can actually \textit{increase} slightly after the matrix refinement procedure.
We have found in our tests that this is the reason for SOFI to set the axis of some symmetry operation slightly different than the other algorithms, and that leads to rare cases when SOFI needs a slightly higher value of symmetry threshold, to obtain the same PG as SYVA or SymMol.
In the set of structures used in our comparison test, the usefulness of the matrix refinement procedure might not be apparent, however when SOFI is used on structures with more disorder in the atomic positions, the refinement procedure is key to the recognition of proper symmetry operations.

The slowest part of the SOFI algorithm is the evaluation of the distance $d_H(A, \boldsymbol{\theta}' A)$ from Eq.~\eqref{eq:dH}, calculated by the CShDA algorithm, which is done for each trial $\boldsymbol{\theta}'$.
It would make sense to spend some computational effort in trying to lower as much as possible the number of times CShDA is computed.
This can be done by recognising when a trial $\boldsymbol{\theta}'$ is very similar to some already known symmetry operation $\boldsymbol{\theta}$, meaning the trial $\boldsymbol{\theta}'$ can be skipped, and thus avoid computing CShDA.
In order to evaluate the similarity between two orthonormal matrices, we use a simple distance function $d(\boldsymbol{\theta}^a, \boldsymbol{\theta}^b)$, which is a sum of the distances between the matrix elements:
\begin{equation}
    d(\boldsymbol{\theta}^a, \boldsymbol{\theta}^b) = \sqrt{\sum_{i,j} (\boldsymbol{\theta}^a_{ij} - \boldsymbol{\theta}^b_{ij})^2}
    \label{eq:matrix_distance}
\end{equation}
which can be seen as a measure of the order of rotation needed to transform $\boldsymbol{\theta}^a$ into $\boldsymbol{\theta}^b$.
By default, SOFI considers two orthonormal matrices, namely $\boldsymbol{\theta}'$ and $\boldsymbol{\theta}$, as equal when the order of rotation needed to transform one into the other is more than 200, that is an operation C200 in the Schoenflies notation, or rotation by angle $0.005\cdot 2 \pi$.
This sets the maximal limit to the resolving power of SOFI, meaning that any symmetry operation of order higher than 200 can not be recognised.
This default behaviour can be changed by tuning of some internal parameters, and recompiling the code. 
The procedure to do that is detailed in the online documentation, available through the front page of the GitHub repository.

With this modification to SOFI, a relatively big speed-up is achieved for structures with a large number of atoms close to the origin (number $N_i$ in Sec.~\ref{sec:sort_atms}), where the redundancy of the search space of matrices $\boldsymbol{\theta}$ is large.
The very extreme example of such structure being the perfect fullerene-type structures,
where all atoms are at the same distance.

The search space of highly symmetric tetrahedral, octahedral, and icosahedral PGs in particular -- $T$, $O$, and $I$ variants -- can be largely redundant. 
In SOFI, this redundancy can be reduced by an early-termination condition, which is triggered when both of the following criteria are met: \textit{i)} either $\sigma$ or $I$ operations are found, and \textit{ii)} there are two or more unique $C_n$ axes with $n \ge 3$. 
The two criteria are sufficient to terminate the search procedure, and directly identify the corresponding PG by the standard procedure.
In principle, this condition can be ambiguous, since a group can be a subgroup of another (for instance, all $T$ variants are subgroups of $Oh$). 
However the ambiguity is lifted by performing the combinations of symmetry elements identified at the current step of the algorithm, which are at that point extremely unlikely to be cyclic, and will thus generate the missing operations.
The overhead of computing the early-termination condition is largely negligible.

The redundancy of the search space however cannot be reduced by the mentioned arguments when the number of symmetry operations is low, as for instance in $C1$, $Cs$, and $Ci$ PGs.
The absolute worst-case performance for the SOFI algorithm as presented in this article is thus a structure with many atoms $N$, all of a single atomic type, where relatively many of them are close to the origin ($N_i$ in Sec.~\ref{sec:sort_atms}), and a low-order PG.

\section{Exemplary uses}
\label{sec:numeric}

\subsection{Generating PG by combination of group elements}
\label{sec:combos}
In SOFI, the assignment of PG name is done by a separate routine, which acts only on the list of operations $\boldsymbol{\theta}$. 
Thus, we can design a small program which reads a number of input operations, generates all possible combinations $\boldsymbol{\theta}'=\boldsymbol{\theta}^i*\boldsymbol{\theta}^j$ in a self-consistent manner, and assigns the PG name to the final list of operations.
In the present example, we use this program to confirm that the D2h and D3d groups are indeed among the subgroups of D6h.
In group theory a high-symmetry PG may lose certain elements to give rise to a new PG, which is a subgroup of the former. Such descent of symmetry can be useful in certain applications. 
The small program of this example is included in the examples directory of the git repository: $\texttt{examples/SOFI/ex\_D6h\_subgroup}$.

The upper part of Table~\ref{tab:D6h} lists the symmetry operations $\boldsymbol{\theta}$ of some atomic structure with the D6h PG (the precise structure is not important for this example). 
If we take only the four operations of labels 5 and 7, and generate all possible unique combinations $\boldsymbol{\theta}'=\boldsymbol{\theta}^i*\boldsymbol{\theta}^j$, we obtain the list of operations in the middle part of Table~\ref{tab:D6h}.
It is straightforward to verify that the obtained list of operations corresponds to D2h PG, and that all the resulting operations are in fact part of D6h, and therefore D2h is a subgroup of D6h.

Similarly, by generating all possible combinations of the four operations labelled 6 and 7 on the upper part of Table~\ref{tab:D6h}, we obtain the list shown in the lower part of Table~\ref{tab:D6h}, which constitutes the D3d PG, and is a subgroup of D6h.
It can be verified that if we take all unique operations from the middle and lower parts of Tables~\ref{tab:D6h}, which is 16 operations, and input them to the combinations program, we obtain back the list of 24 operations of D6h from the upper section of Table~\ref{tab:D6h}.

\begin{table}[h]
\centering
    \footnotesize
    \begin{NiceTabular}{c|c|c r r r}
   &      Label & Operations $\boldsymbol{\theta}$ & & Axis & \\ \hline
\Block{8-1}{\rot{D6h}}  
   &  0   &  $E$, $I$  & & &   \\
   &  1   &  $\sigma$, $C_2$, $C_3$, $C_6$, $S_3$, $S_6$ &  0.989 &  -0.094 &  -0.109   \\
   &  2   &  $\sigma$, $C_2$ &   -0.052 &  -0.939 &  0.339    \\
   &  3   &  $\sigma$, $C_2$ &   -0.134 &  -0.330 & -0.934    \\
   &  4   &  $\sigma$, $C_2$ &   -0.022 &   0.648 & -0.761    \\
   &  5   &  $\sigma$, $C_2$ &   -0.112 &  -0.978 & -0.173    \\
   &  6   &  $\sigma$, $C_2$ &   -0.142 &  -0.756 & -0.639    \\
   &  7   &  $\sigma$, $C_2$ &    0.090 &  -0.183 &  0.979    \\ \hline
\Block{4-1}{\rot{D2h}}
   &  0 & $E$, $I$ & & &  \\ 
   &  1 & $\sigma$, $C_2$  &    -0.112 &  -0.978 &  -0.173  \\
   &  2 & $\sigma$, $C_2$  &     0.090 &  -0.183 &   0.979  \\
   &  3 & $\sigma$, $C_2$  &     0.989 &  -0.094 &  -0.109  \\ \hline
\Block{5-1}{\rot{D3d}}
   &  0  & $E$, $I$ & & & \\
   &  1  &  $C_3$, $\overline{C_3}$, $S_6$, $\overline{S_6}$  &  0.989 &  -0.094 &  -0.109  \\
   &  2   &  $\sigma$, $C_2$  &   0.090  &  -0.183 &  0.979  \\
   &  3   &  $\sigma$, $C_2$   &  -0.142 &  -0.756 & -0.639  \\
   &  4   &  $\sigma$, $C_2$   &  -0.052 &  -0.939 &  0.339  \\
    \end{NiceTabular}
    \caption{
    Upper part: list of operations $\boldsymbol{\theta}$ of some structure with the D6h PG, grouped together by the axis they act on. The transposes of applicable operations ($\overline{C_3}$, $\overline{C_6}$, $\overline{S_3}$, $\overline{S_6}$) on axis 1 are omitted for clarity. \\
    Middle part: the resulting operations of all possible combinations of four operations on axes 5 and 7 from the upper part. The PG with these operations is D2h, and all the operations are also included in D6h. \\
    Lower part: the resulting operations of all possible combinations of four operations on axes 6 and 7 from the upper part. 
    The PG with these operations is D3d, and all the elements are also included in D6h.
    }
    \label{tab:D6h}
\end{table}

\subsection{Obtaining symmetry-equivalent configurations}
\label{sec:Pt111}
\begin{figure}[h]
    \includegraphics[width=\columnwidth]{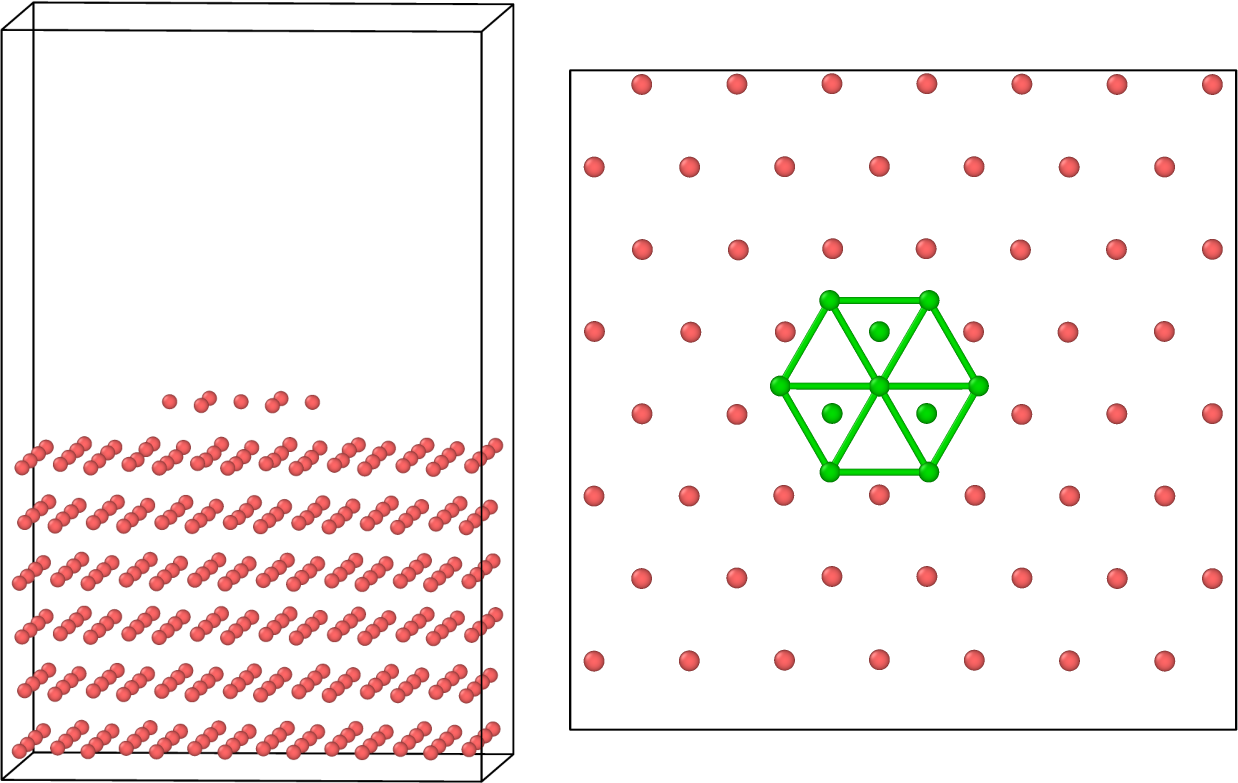}
    \caption{Initial state configuration of the heptamer island on Pt(111) surface. Left: side view. Right: top view of the surface layer, with bonds drawn between the atoms of the island. Marked in green are the atoms of the local configuration that is used as input for SOFI.}
    \label{fig:pt111}
\end{figure}

The power of knowing the set of local symmetries $\boldsymbol{\theta}$
is demonstrated on the example system of heptamer island on Pt(111) surface, taken from the OptBench database~\cite{OptBench}, shown on Fig.~\ref{fig:pt111}. 

\begin{figure}[bht]
  \includegraphics[width=\columnwidth]{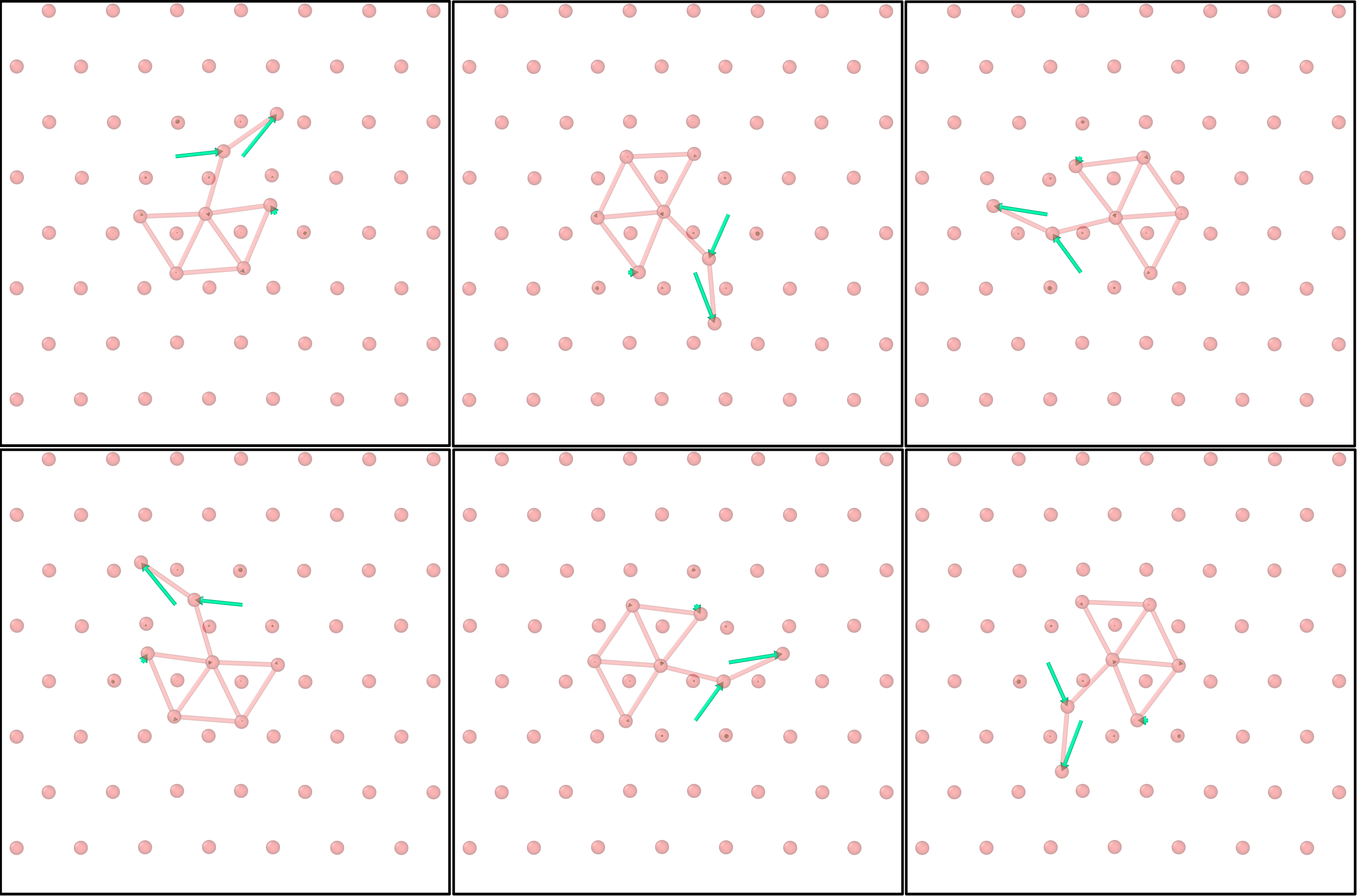}
  \caption{First panel (top left) shows $\Delta_1$ in cyan, which is on other panels transformed by the set \{$\boldsymbol{\theta}_j$, $P_j$\}, and represents saddle points of the PES equivalent by local symmetry. The top row contains the rotations of $\Delta_1$ about the axis through the origin atom (out of plane axis), and the bottom row contains reflections of $\Delta_1$ over three mirror planes.}
  \label{fig:saddles}
\end{figure}

A particular, known displacement of the atoms from the initial structure is shown on Fig.~\ref{fig:saddles} in six symmetry-equivalent variants in each panel, with the displacement vectors marked in cyan, each labelled $\Delta_j$, for $j\in [1,6]$.
The displacement of the atoms by specifically these vectors actually corresponds to equivalent configurations at the saddle point of the potential energy surface (PES) of the system, the details of which are not important for the purpose of this example.
By knowing only one of these displacements, say $\Delta_1$ shown on the top left panel of Fig.~\ref{fig:saddles}, the other five can be generated by only using the information of local symmetry of the heptamer island, which can be computed by SOFI.

In the present example, this was done by first selecting the first-neighbour local configuration around the atom at the center of the heptamer, shown in green on Fig.~\ref{fig:pt111}, where the set of atomic indices contained in the local configuration is labelled $[s]$, and the atom at the center of the heptamer is taken as the origin point.
This local configuration was then input to SOFI as an independent atomic structure, which returns the set of six symmetry operation matrices and the corresponding permutations \{$\boldsymbol{\theta}_j$, $P_j$\}, in relation to the desired origin. The symmetry operations are three rotations about the vertical axis through the origin atom, and three mirror reflections.
The displacement $\Delta_1[s]$ on the subset of atoms $[s]$ can then be transformed by applying each matrix $\boldsymbol{\theta}_j$, followed by permutation of the vector elements $P_j$, to obtain the other five displacements $\Delta_j [s]$,
\begin{equation}
    \Delta_j [s] = P_j \big(~\boldsymbol{\theta}_j \Delta_1[s]~\big)
\end{equation}
which correspond to the displacement vectors of Fig.~\ref{fig:saddles}, and bring the system to the symmetry-equivalent saddle point configurations.

The example shown in this section illustrates a procedure which exploits the list of symmetry operations, and applies them on some property of the atomic structure to obtain the symmetry-equivalent representations of that property, in this case a displacement.
By analogy, such procedure can be done to any vectorial property of a structure, for which symmetry-equivalent representations exist.
The symmetrization of an atomic structure is an example of such procedure (with some modifications), acting on the atomic positions.



\section{Conclusion}
\label{sec:conclusion}
We have developed the SOFI algorithm for finding PG symmetry operations of non-periodic atomic structures, based on the IRA shape matching algorithm~\cite{ira}.
The working principle of SOFI is to cast the problem of finding PGs as a shape-matching problem of a single structure with itself, and find all the degenerate solutions.
The set of these solutions corresponds to the symmetry operations of the atomic structure.

Contrary to some other algorithms, SOFI is not limited to the number of atoms that a structure can contain.
It is also agnostic to the choice of the origin point, which allows the identification of symmetry operations about any chosen point/atom, not only the geometric center.

We have tested SOFI on a large set of atomic structures where symmetry identification can be challenging, such as single-element clusters and close-packed structures. 
The former are critical because of the large number of possible permutations of atomic indices, and the latter due to the rather homogeneous spatial distribution of atoms, which prevents simple identification of any principal axes.
Both these cases are of relevance for numerical applications, for instance in processing substructures obtained from other condensed phase (periodic) simulations, whenever a local symmetry assignment is required for clusters or slabs, and hence where an automatic assignment of symmetry is highly desirable. 
Due to the way SOFI parses through the space of possible operations, it can be advantageous in such critical cases.

We have compared the results of our SOFI algorithm to three other algorithms for PG identification from the literature, namely SYVA, SymMol, and libmsym, at two different values of the symmetry threshold.
The results are given in a statistical manner that shows the general behaviour of SOFI, as compared to other algorithms.
SOFI was able to successfully identify PGs across all sets, with a large degree of agreement with other algorithms. 
In most cases of disagreement between algorithms, SOFI has found a higher number of symmetry elements within the specified symmetry threshold, and thus a higher order point group. 
In all cases where SOFI found a lower order PG than another algorithm, we have been able to retrieve the higher order PG by tuning of the symmetry thresholds parameter of SOFI.

We also present two examples of how SOFI can be utilised, other than simple identification of PGs; in the first example as a tool to generate combinations of several matrices until group completeness; and in the second example as tool to aid automatic generation of structures, and their properties (in this case a displacement vector), that transform according to the local PG symmetry about a specific atom, \textit{i.e.} any covariant property.

One of the core ideas of SOFI is to provide a modular, and interoperable set of computational routines, which could be used for retrieving the point group symmetry properties of a structure, as standalone, or within another application.
The inclusion of SOFI into the IRA library makes it a convenient, and accessible tool for purposes related to the precise analysis of atomic structures.

\section*{Supplementary Materials}
The supplementary material contains all results of the comparison tests done in this work. The results are reported in tables and images, where the tables report the result (PG) of each algorithm included in the test for each structure, and the images serve as an overall comparison between the algorithms.


\begin{acknowledgements}
    The authors are active members of the Multiscale And Multi-Model ApproacheS for Materials In Applied Science consortium (MAMMASMIAS consortium), and acknowledge the efforts of the consortium in fostering scientific collaboration. 
    The work was partly funded by the Croatian Science Foundation (HrZZ), via the project IP-2020-02-7262 (HYMO4EXNOMOMA).
    All images of atomic structures in this article were generated with Ovito~\cite{ovito} software.
\end{acknowledgements}

\section*{Author declarations}
\subsection*{Conflict of interest}
The authors have no conflicts to disclose.

\subsection*{Author contributions}
\noindent
\textbf{M. Gunde:} Conceptualization, Data curation, Software, Writing - original draft, Writing - review \& editing, Supervision, Visualization;
\textbf{N. Salles:} Conceptualization, Writing - review \& editing, Supervision, Visualization;
\textbf{L. Grisanti:} Conceptualization, Writing - review \& editing, Supervision, Visualization;
\textbf{L. Martin-Samos:} Conceptualization, Writing - review \& editing, Supervision, Visualization;
\textbf{A. Hemeryck:} Conceptualization, Writing - review \& editing, Supervision, Visualization;

\section*{Data and source code availability}
SOFI is released as part of the IRA library, which is under double licensing, GPL v3 and Apache v2. 
The source code and example programs are available at \url{https://github.com/mammasmias/IterativeRotationsAssignments}.
Link to the online documentation is available from the main GitHub page, which includes some tutorials, and detailed descriptions of the source code, the C-bound API, and the python interface.
All data used in the comparison tests is openly available from the CCD~\cite{cdb}.
All datasets and scripts used in the comparison test, and plain-text results are available from Zenodo~\cite{tests_zenodo}.

\appendix

\section{Some properties of PG symmetry matrices}
\label{app:matrices}
The matrices $\boldsymbol{\theta}$ corresponding to PG symmetry operations are orthonormal ($\boldsymbol{\theta}^{-1}=\boldsymbol{\theta}^T$), non-symmetric matrices with real elements, and correspond to either pure rotations (symbol $C$), pure reflections (symbol $\sigma$), or combinations of both (rotoreflections, symbol $S$). 

The matrices $\boldsymbol{\theta}_R$ corresponding to pure rotations (symbol $C$) have a positive determinant, $det(\boldsymbol{\theta}_R)=1$, the value of trace is $Tr(\boldsymbol{\theta_R})=1+2\cos(\alpha)$, and their eigenvalues are $\lambda_1 = 1$, and $\lambda_{2,3}=\exp{(\pm i \alpha)}$ where $\alpha$ is the angle of rotation about the axis given by the eigenvector corresponding to $\lambda_1$.
A pure rotation matrix can be constructed from knowing the angle and axis of rotation, for example by the well-known Euler-Rodrigues rotation formula.

Matrices $\boldsymbol{\theta}_F$ which represent pure reflection (symbol $\sigma$) are characterised by the negative determinant, $det(\boldsymbol{\theta}_F)=-1$, the value of trace $Tr(\boldsymbol{\theta}_F)=1$, and their eigenvalues are $\lambda_1=-1$, and $\lambda_{2,3}=1$. 
The normal vector of the plane of reflection is given by the eigenvector of $\boldsymbol{\theta}_F$ corresponding to $\lambda_1$.
A reflection matrix can be constructed from knowing the normal vector of the plane of reflection $\ket{x}$ by $\boldsymbol{\theta}_F=E - 2\ket{x}\bra{x}$, where $E$ is the identity matrix, and $\ket{.}\bra{.}$ indicates the outer product.

Matrices $\boldsymbol{\theta}_{RF}$ corresponding to rotoreflections (symbol $S$) are characterised by the negative determinant, $det(\boldsymbol{\theta}_{RF})=-1$, the value of trace $Tr(\boldsymbol{\theta}_{RF})=-1+2\cos(\alpha)$, and their eigenvalues are $\lambda_1=-1$, and $\lambda_{2,3}=\exp{(\pm i\alpha)}$, where $\alpha$ is the angle of rotation about the axis given by the eigenvector of $\lambda_1$, which is at the same time also the normal vector of the plane of reflection.
Matrices corresponding to rotoreflections can be constructed from combination of pure rotation and pure reflection matrices, $\boldsymbol{\theta}_{RF} = \boldsymbol{\theta}_F \boldsymbol{\theta}_R$, where $\boldsymbol{\theta}_F$ and $\boldsymbol{\theta}_R$ have the same vector as the axis of rotation, and the normal of the plane of reflection.

The angle $\alpha$ in radians is transformed to integers $n$ and $p$, such that ${\alpha}/{2\pi} = {p}/{n}$. 
For instance, the rotation for 5/12th of a circle (angle 150$^{\circ}$), would be $n=12$ and $p=5$.

It can be observed that the use of $\sigma$ symbol for pure reflections over some axis is redundant, since an equivalent notation is a rotoreflection $S$ with angle $\alpha=0$. We thus label pure reflections as "$S~0$".

The properties listed above are synthesised in the Table~\ref{tab:mat_prop}.

\begin{table}[htb]
    \centering
    \begin{tabular}{c||c| c |c| c}
         &  identity $E$ & inversion $I$ & rotation $C$ & rotoreflection $S$ \\ \hline
     $det(\boldsymbol{\theta})$    & 1 & -1 & 1 & -1 \\ \hline
     $\lambda_{1,2,3}$ & 1, 1, 1 & -1, -1, -1 & 1, $\exp{(\pm i\alpha)}$ & $-1$, $\exp{(\pm i\alpha)}$ \\\hline
     $Tr(\boldsymbol{\theta})$ & 3 & -3 & $1+2\cos(\alpha)$ & $-1+2\cos(\alpha)$
    \end{tabular}
    \caption{The axis of rotation (plane of reflection) is given by the eigenvector corresponding to $\lambda_1$.}
    \label{tab:mat_prop}
\end{table}

The slight exceptions to the above are the identity $E = diag(1,1,1)$ and the point-inversion $I=diag(-1,-1,-1)$, which have the same matrix form regardless of the orientation of the reference frame.
Nevertheless, the identity matrix can be seen as rotation about any axis for an angle $\alpha=0$ (operation $C~0$), and the point-inversion can be seen as rotoreflection about any axis for an angle $\alpha=\pi$ (operation $S~2^{\wedge}1$).

Due to the numerical procedure of matrix diagonalization, where any $\pm$ direction is an equivalent solution for the eigenvectors (thus the axis), the relative orientation of the operation (sign of the angle $\alpha$) cannot be trusted.
For this reason, we follow the convention for orienting the axis such that its components are: $z > 0$, if $z=0$ then $x>0$, and if $x=0$ then $y>0$. 
The sign of $\alpha$ is then deduced with respect to this axis.

\bibliography{biblio}

\end{document}